\documentclass[12pt]{article}
\input psfig.tex

\begin{document}
\begin{center}
{\bf Large-Scale Magnetic Field Re-generation by Resonant MHD Wave Interactions} \\

S. Galtier\\

Institut d'Astrophysique Spatiale (IAS), B\^atiment 121, F-91405 Orsay 
(France); Universit\'e Paris-Sud 11 and CNRS (UMR 8617)\\

and\\

S. Nazarenko\\

Mathematics Institute, The University of Warwick, Coventry, CV4-7AL, UK\\

\end{center}
\vskip0.5truein

\begin{abstract}
We investigate numerically the long-time behavior of balanced Alfv\'en wave turbulence forced 
at intermediate scales. Whereas the usual constant-flux solution is found at the smallest scales, 
two new scalings are obtained  at the forcing scales and at 
the largest scales of the system. In the latter case we show, in particular, that the spectrum 
evolves first to a state determined by  Loitsyansky invariant and later a state close to the 
thermodynamic equipartition solution predicted by wave turbulence. The astrophysical implications for 
galactic magnetic field generation are discussed. 
\end{abstract}

\section{Introduction}

Turbulence flows are ubiquitous in astrophysical environments from the solar wind 
(Goldstein et al., 1999; Galtier, 2006), to interstellar (Elergreen et al., 2004; Scalo et al., 2004), 
galactic and even 
intergalactic media (Govoni et al., 2006). At the larger scales, signatures of astrophysical turbulence are 
found, in particular, in the magnetic field measurements whose origin remains one of the major 
challenging problems (Pouquet, 1993; Widrow, 2002; Brandenburg, 2005). 
In this paper, we emphasize a new mechanism for generating large-scale magnetic field. It is based 
on the resonant interactions of shear-Alfv\'en waves in a turbulent medium permeated by a strong 
external magnetic field and influenced by an external forcing at intermediate scales. In this situation, 
both large-scale kinetic and magnetic energies may be produced by mainly non local interactions. 
This scenario, although very simple, may be relevant for describing  re-generation (i.e. maintenance) of
large-scale galactic fields, e.g. in our galaxy where energy is injected at intermediate 
scales by stellar winds and supernovae explosions on scales $10-100$pc 
(Ferriere et al., 2004).

\section{Alfv\'enic turbulence}

The wave kinetic equations of incompressible MHD were derived rigorously by Galtier et al.
(2000,2002). Let us consider
the simplest case when helicities are taken to be zero, when pseudo-Alfv\'en 
waves are discarded and when turbulence is axially symmetric and 
balanced (equal spectra for Alfv\'en waves co- and counter-propagating
with respect to the external magnetic field).  In this case for the 
shear-Alfv\'en wave energy spectrum we have 
$ E(k_\perp,k_\parallel) = f(k_\parallel) E_\perp(k_\perp), $
where $f(k_\parallel)$ is a function fixed by the initial conditions (i.e. there
is no energy transfer in the parallel direction), and the transverse
part obeys the following kinetic equation, (Galtier et al., 2000)
\begin{eqnarray}
\partial_t E_\perp(k_\perp) + \nu k_{\perp}^2 E_\perp(k_\perp) = 
\frac{\pi}{b_0} \int \cos^2\phi \sin\theta  \, \frac{k_\perp}{\kappa_\perp}
E_\perp(\kappa_\perp) \nonumber \\
\left[k_\perp E_\perp(L_\perp)-L_\perp E_\perp(k_\perp)\right]
d\kappa_\perp dL_\perp . 
\label{kineticE}
\end{eqnarray}
Here $b_0$ the background magnetic field (which 
will be taken equal to one), $\phi$ the angle between $\mathbf{k_\perp}$ and $\mathbf{L_\perp}$, 
and $\theta$ the angle between $\mathbf{k_\perp}$ and $\mathbf{\kappa_\perp}$. This 
integro-differential equation has been computed in (Galtier et al., 2000) but only for short times in order to 
analysis, in particular, the formation of the Kolmogorv-Zakharov solution. In this paper we investigate 
the long-time behavior of Alfv\'en wave turbulence when of a forcing term is applied. In principle, 
this regime may be investigated for any value ($\not= 1$) of the magnetic Prandtl number; in this 
case the viscosity $\nu$ has to be seen as a combination of the kinematic viscosity and the 
magnetic diffusivity. Note that in the alfv\'enic turbulence regime, the kinetic and magnetic energies 
are equal, therefore all results may be interpreted in terms of magnetic energy fluctuations.

\subsection{Short-time evolution}

We perform numerical simulations of the wave kinetic equation (\ref{kineticE}) of MHD turbulence 
in which a 
forcing term is introduced at intermediate scales. A non-uniform grid in Fourier space is used 
(Galtier et al., 2000) with $k_{\perp}(i) = \delta k \ 2^{i/F}$ ($i \in [0, 156]$; $\delta k=1/8$; $F=8$). This 
method allows us to achieve very high Reynolds numbers ($>10^5$ with $\nu=\eta=5 \times 
10^{-3}$). A weak forcing  is applied to the flow at wavenumbers $k_{\perp} \in 
[2^{-1/8},2^{19/4}]$ with a constant (flat) spectrum. 

\begin{figure}
\centerline{\psfig{figure=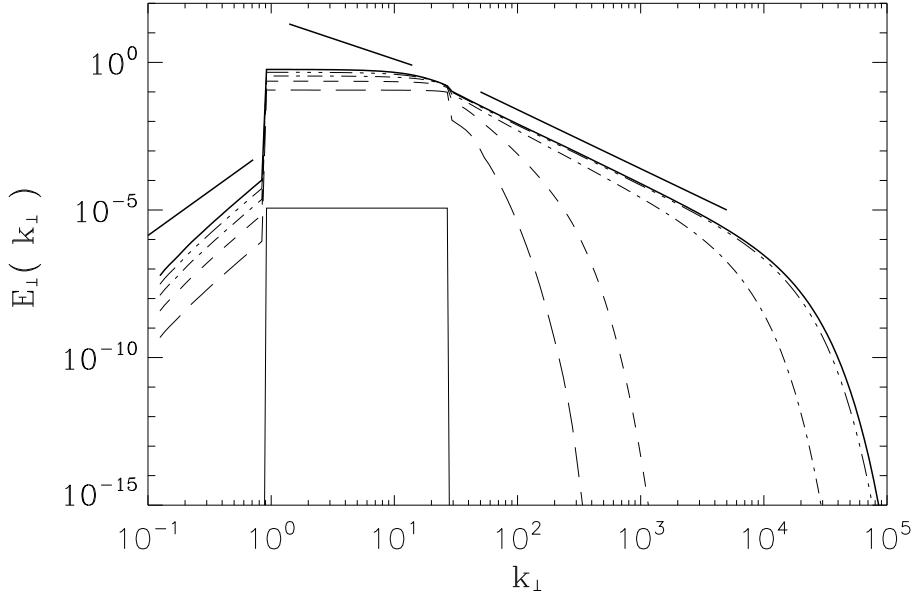}}
\caption{Evolution of the energy spectrum for times $t_0=0$ (solid),
$t_1=5\times 10^{-4}$ (long dash), $t_2=2t_1$ (small dash), $t_3=3t_1$ (dot-dash),
$t_4=4t_1$ (three dots-dash) and $t_5=5t_1$ (bold 
solid). For comparison, three slopes are given with power law indices $+3$, $-1.4$ and $-2$ 
respectively.} 
\label{Fig1}
\end{figure}
In Figure \ref{Fig1} the evolution of the shear-Alfv\'en wave energy spectrum is displayed for 
short times. The forcing scales  determine the initial evolution as can be
 seen at intermediate 
wavenumbers. 
The first robust property to be observed is the generation of small-scales through a direct cascade and the 
formation of the well-know $k_{\perp}^{-2}$ solution predicted by wave turbulence theory 
(Galtier et al., 2000). At the same time we see that the largest scales of the MHD flow are affected by the 
forcing and produce a power law slightly steeper than the reference slope ($+3$). At intermediate 
scales, no clear tendency is observed yet, only a reduction of the flat spectrum seems to occur. 

\begin{figure}
\centerline{\psfig{figure=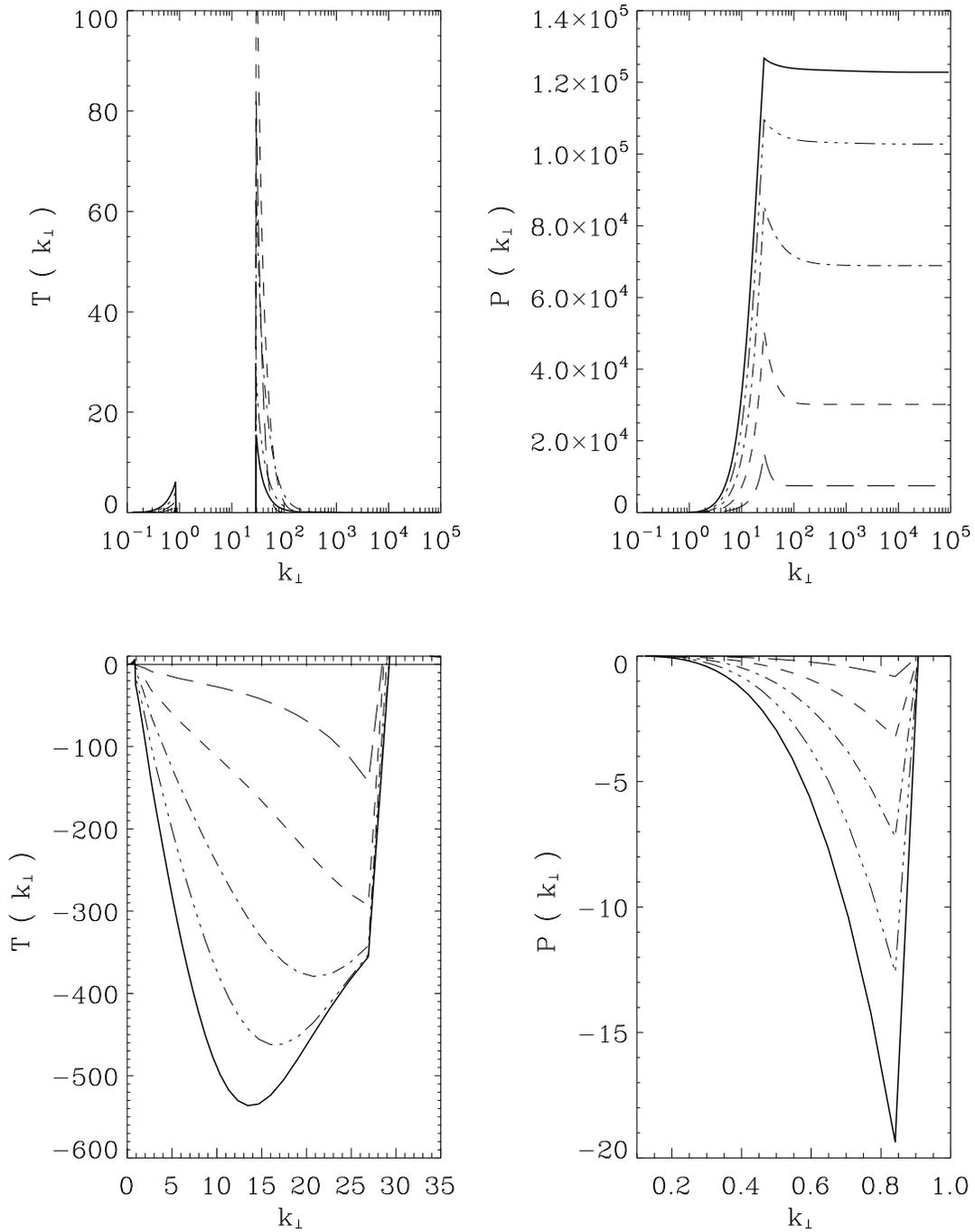}} 
\caption{Temporal evolution of the transfer function $T(k_{\perp})$ and the flux $P(k_{\perp})$ 
at the same times as in Figure \ref{Fig1} (the same notation are also used). Top: general behavior 
at all scales of $T$ and $P$; note the formation of the constant flux solution at small-scales. 
Bottom: zoom at large-scales; a negative transfer is clearly seen at intermediate (forcing) scales 
whereas a negative flux is found at the largest scales.}
\label{Fig2}
\end{figure}

In Figure \ref{Fig2}, we show the temporal evolution of the transfer function, $T(k_{\perp})$, and 
the energy flux, $P(k_{\perp})$, for the same times as in Figure \ref{Fig1}. We remind the general 
relation in the inertial range (where forcing and dissipation are negligible) 
\begin{eqnarray}
\partial_t E_\perp = T = -\frac{\partial P}{\partial k_\perp} .
\end{eqnarray}
The constant flux solution of wave
turbulence is clearly observed at small-scales and corresponds to a direct cascade (positive
flux). As expected, a negative transfer is found at the forcing scales but a strong asymmetric 
profile is obtained initially which reduces with time. This reduction of asymmetry appears when 
the $k_{\perp}^{-2}$ solution is well established (see Figure \ref{Fig1}). 
At the largest scales a positive transfer with a negative (non constant) flux are found which 
means that energy is accumulated probably mostly by non-local interactions. 
Note that  this is not an inverse cascade behaviour because 
there is no extra (in addition to the energy) positive invariant in our problem
that could lead to a dual cascade process.

\subsection{Intermediate-time evolution}

\begin{figure}
\centerline{\psfig{figure=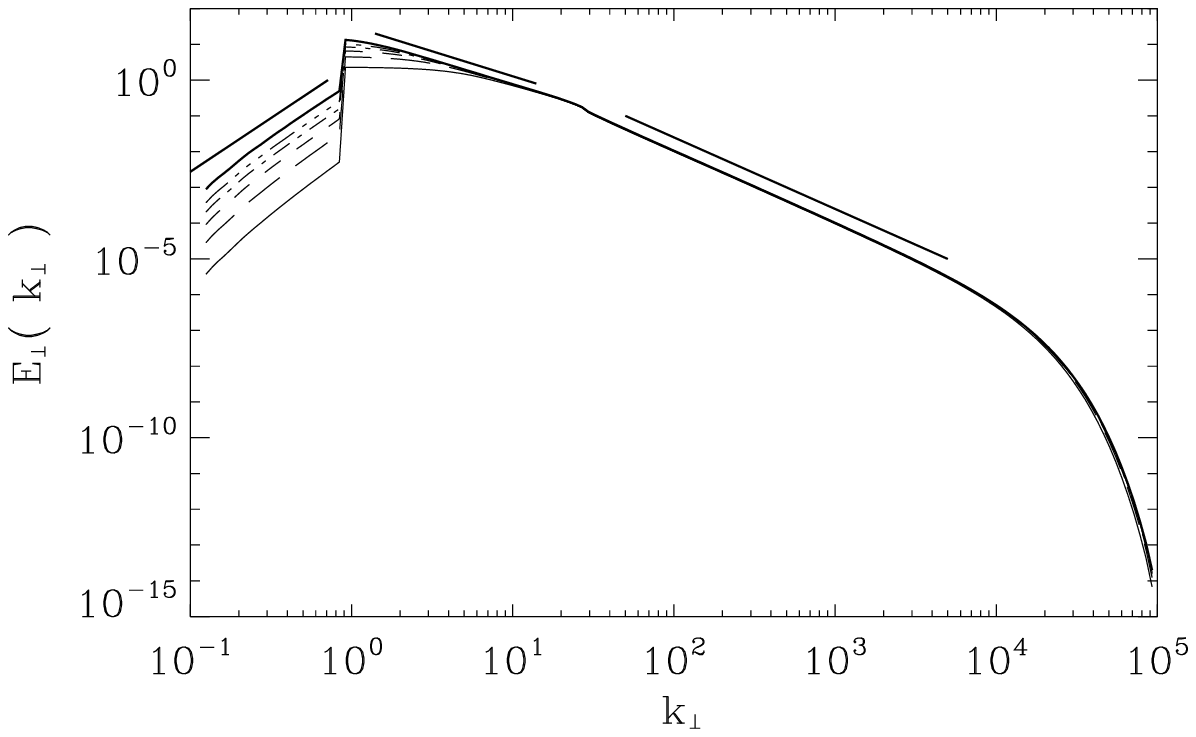}} 
\caption{Evolution of the energy spectrum for times $t_6=0.75\times 10^{-2}$ (solid), 
$t_7=1.75\times 10^{-2}$ (long dash), $t_8=2.75\times 10^{-2}$ (small dash), 
$t_9=3.75\times 10^{-2}$ (dot-dash), $t_{10}=4.75\times 10^{-2}$ (three dots-dash) and 
$t_{11}=0.1$ (bold solid). The same slopes as in Figure \ref{Fig1} are given.} 
\label{Fig3}
\end{figure}

In Figure \ref{Fig3}, we show a next set of times that we call intermediate-time. Whereas the 
small-scale power-law spectrum appears to be stable with no apparent change, at the largest scales 
a $k_{\perp}^{3}$ scaling is found. This solution could be related with the invariance of 
the Loitsyansky integral (Davidson, 2004; Bigot et al., 2007) which leads to scaling for the modal spectrum in 
$k^{s-2}$ at low wavenumbers with $s=D+1$ ($D$ being the space dimension). This analysis 
is often advocated to justify the scaling taken initially in (direct) numerical simulations. For example, 
it is shown in a recent work (Bigot et al., 2007) that the $k_{\perp}^{3}$ scaling is well conserved in freely 
decaying wave MHD turbulence. 
\begin{figure}
\centerline{\psfig{figure=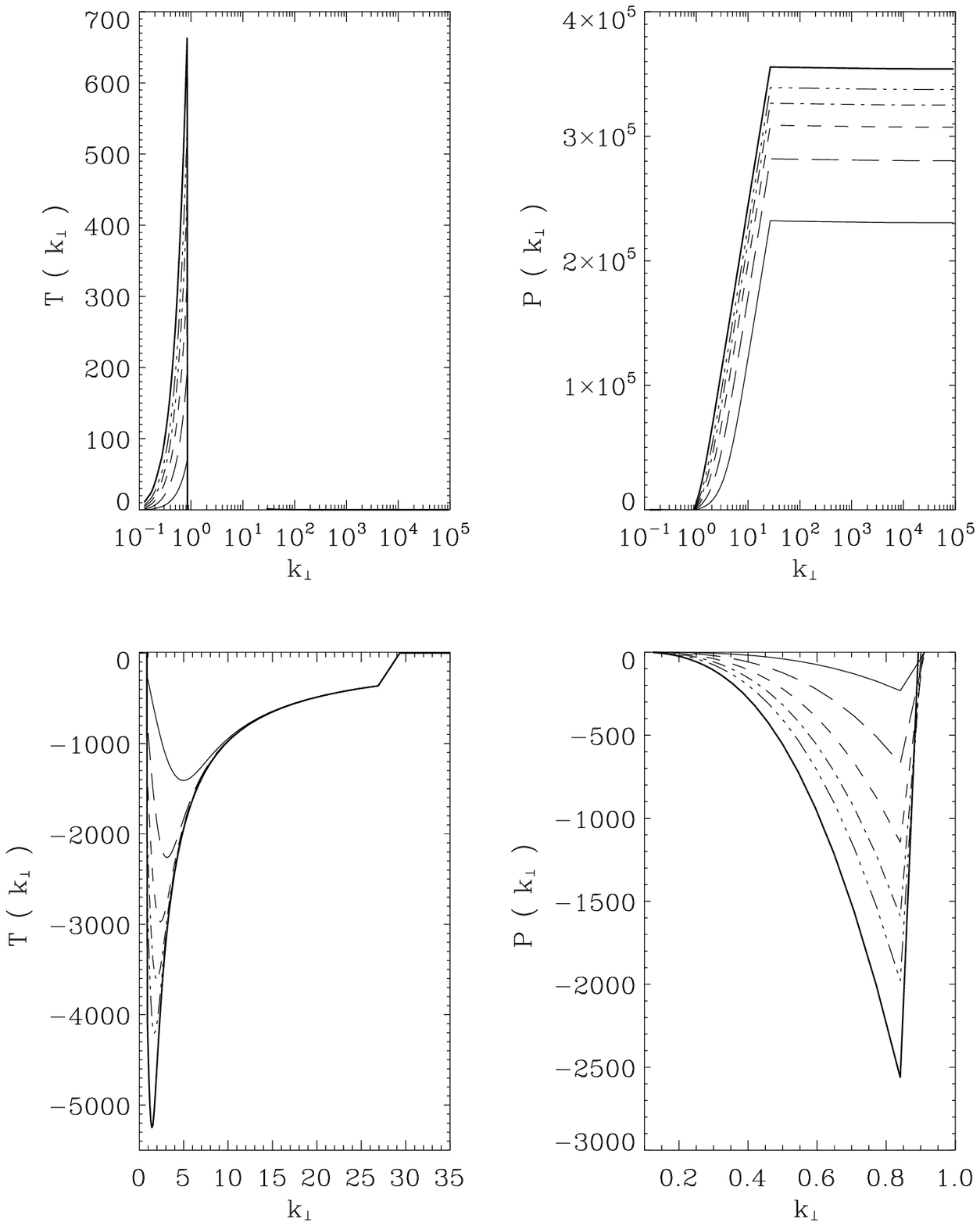}} 
\caption{Temporal evolution of the transfer function $T(k_{\perp})$ and the flux $P(k_{\perp})$ 
at the same times as in Figure \ref{Fig3}. Top: general behavior at all scales. Bottom: zoom at 
large-scale.} 
\label{Fig4}
\end{figure}
At the  intermediate scales, we see clearly a new scaling close to $k_{\perp}^{-1.4}$. It takes about 
one order of magnitude longer to build this solution than to form the small-scale
cascade spectrum. We will see that this solution may be explained 
rigorously by an analysis based on the wave kinetic equation. 

Figure \ref{Fig4} gives the transfer function and flux for the same times as in Figure \ref{Fig3}. 
The constant flux at small-scale is still observed with an increase of its value in time; note 
that this increase seems to saturate.  At wavenumbers around one, we see that the transfer function 
exhibits a change of sign going from a positive value, for wavenumbers smaller than one, to a negative 
value, for wavenumbers larger than one, which traduces an energy transfer towards larger scales. 
In the meantime, we note at the largest scales a decreasing of the flux which is still negative.

\subsection{Asymptotically long-time evolution}

\begin{figure}
\centerline{\psfig{figure=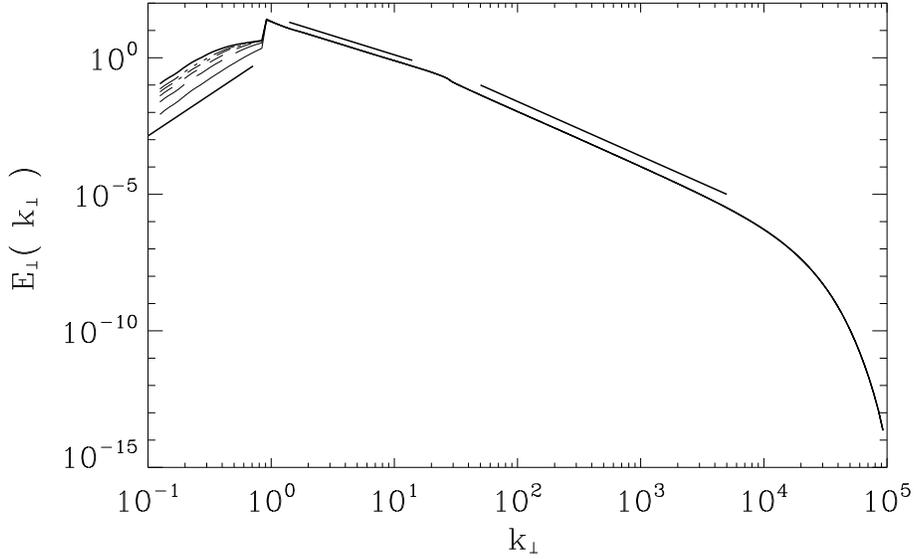}} 
\caption{Long-time evolution of the energy spectrum at $t_{12}=0.15$ (solid), $t_{13}=0.35$ (long dash), 
$t_{14}=0.55$ (small dash), $t_{15}=0.75$ (dot-dash), $t_{16}=0.95$ (three dots-dash) and 
$t_{17}=1.45$ (bold solid).} 
\label{Fig5}
\end{figure}

The long-time behavior of Alfv\'en wave turbulence has been investigated as well. Figure \ref{Fig5} 
shows the energy spectrum evolution for times up to $t=1.45$. Only the large-scale flow evolves  at this
stage, namely
from the $k_{\perp}^{3}$ scaling to a flatter power law. 

\begin{figure}
\centerline{\psfig{figure=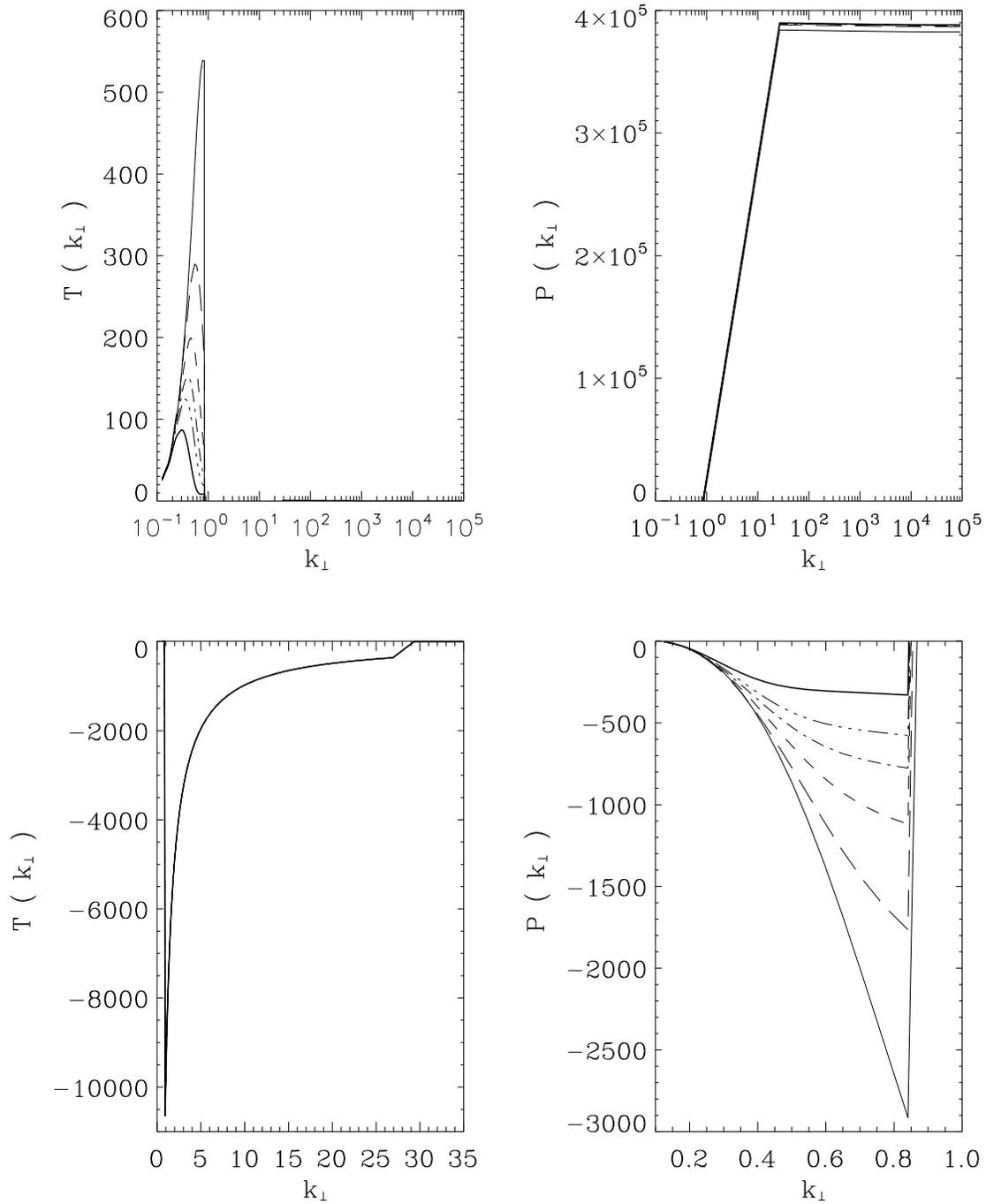}} 
\caption{Temporal evolutions of the transfer function $T(k_{\perp})$ and the flux $P(k_{\perp})$ 
at the same times as in Figure \ref{Fig5}.} 
\label{Fig6}
\end{figure}
Figure \ref{Fig6} gives more interesting information. 
The tendency to flux saturation observed at the smallest scales is confirmed and the transfer 
function at intermediates scale does not change anymore. As expected, only the largest scales 
are evolving with a transient  decrease of the absolute value of the flux (which is negative at
the large scales) accompanied by a plateau formation and a 
decrease of the transfer function. 

\begin{figure}
\centerline{\psfig{figure=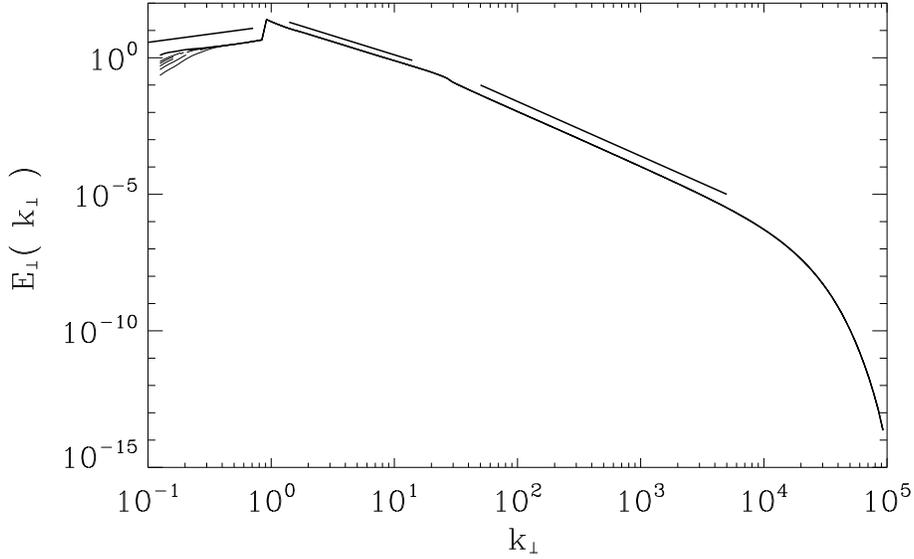}} 
\caption{Asymptotically long-time evolution of energy spectrum at times $t_{18}=2.5$ (solid), 
$t_{19}=4.5$ (long dash), $t_{20}=6.5$ (small dash), $t_{21}=8.5$ (dot-dash), $t_{22}=10.5$
(three dots-dash) and $t_{23}=24.5$ (bold solid).} 
\label{Fig7}
\end{figure}

The last set of graphs displays the final behavior (up to time $t=24.5$) of Alfv\'en wave 
turbulence. Figure \ref{Fig7}  
reveals the formation of a new power law close to $k_{\perp}^{0.6}$ at the largest scales whereas 
the other scales are still steady. 

\begin{figure}
\centerline{\psfig{figure=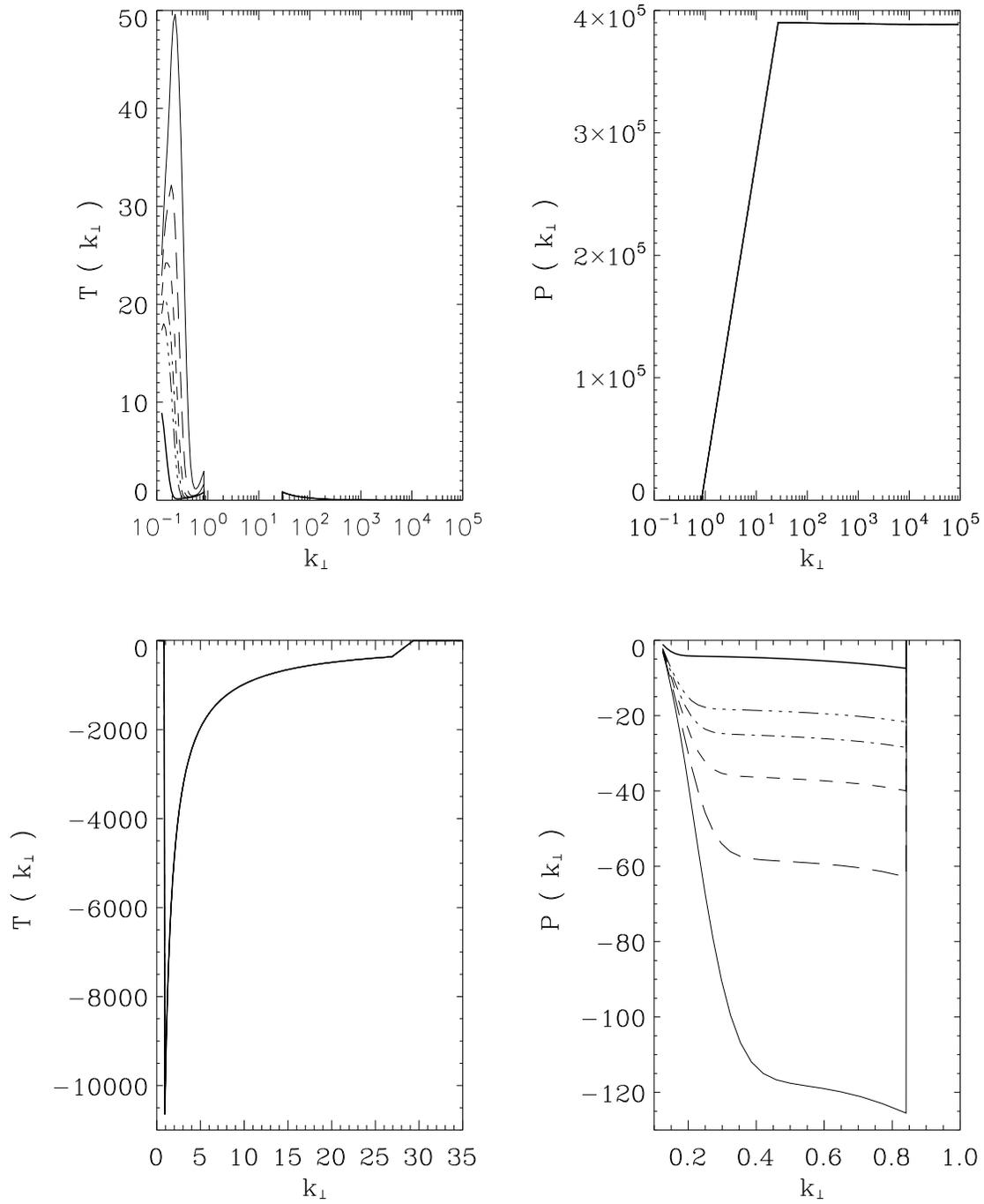}} 
\caption{Temporal evolutions of the transfer function $T(k_{\perp})$ and the flux $P(k_{\perp})$ 
at the same times as in Figure \ref{Fig7}.}
\label{Fig8}
\end{figure}
In Figure \ref{Fig8} the flux weakening and the plateau formation is continuing
at the largest scales   whereas the transfer function is still decreasing.

\section{Discussion}
\subsection{Nonlinear dynamics at intermediate (forcing) scales}

The scaling behavior found at intermediate scales seems to be strongly dependent on the 
type of forcing applied to the MHD flow. This is confirmed with a second simulation (not shown) 
where a non flat forcing is applied. The nonlinear dynamics at forcing scales may be investigated 
through a rigorous analysis based on the wave kinetic equations. Let us start from equation 
(\ref{kineticE}) and assume a power law scaling solution
\begin{eqnarray}
E_\perp(k_\perp) = C k_{\perp}^n . 
\end{eqnarray}
Then let us assume the following general forcing 
\begin{eqnarray}
f(k_\perp) = C_f k_{\perp}^{\xi} . 
\end{eqnarray}
Therefore, we have 
\begin{eqnarray}
\partial_t E_\perp(k_\perp) + \nu k_{\perp}^2 E_\perp(k_\perp) = C.I. + f(k_\perp) ,
\end{eqnarray}
where C.I. is the collision integral. We focus our analysis on the forcing scales and neglect the 
dissipative term. By introducing the notation $\kappa_\perp = k_\perp x$ and $L_\perp = 
k_\perp y$, we obtain the steady solution
\begin{eqnarray}
C^2 k_\perp^{2n+3} I(n) \sim C_f k_{\perp}^{\xi} , 
\end{eqnarray}
where the collision integral writes
\begin{eqnarray}
I(n) = \int_{\Delta} \left(\frac{y^2+1-x^2}{2y}\right)^2 \sqrt{1- \left(\frac{x^2+1-y^2}{2x}\right)^2} 
x^{n-1} y (y^{n-1}-1) dx dy .
\end{eqnarray}
Since $I(n)$ is not singular, we directly obtain the scaling relation 
\begin{eqnarray}
n = \frac{\xi-3}{2} . 
\end{eqnarray}
For a constant (flat spectrum) forcing, we get $n=-3/2$. The solution found is indeed very 
close to this power law solution. The slight difference observed may be attributed to the non 
exact stationarity of the solution (see Figure \ref{Fig9}). Note that we have tested the 
theoretical prediction against another forcing with $\xi=1$. We find a steady energy spectrum 
in $k_\perp^{-0.9}$ at intermediate scales which is close to the prediction in $k_\perp^{-1}$ 
whereas the large-scale spectrum seems not to be affected and behaves similarly as in the 
simulation presented in this paper.

\subsection{Solution at the largest scales}

The asymptotically long-time evolution is characterized at the largest scales by a decreasing
and  flat in $k$ energy flux 
 (see Figure \ref{Fig8}). Contrary to what happens at the smallest scales, the flux has 
a very small value (around $-5$ compare to $4 \times 10^5$ at small scales) whereas the transfer 
function tends to zero. Therefore, the final state of Alfv\'en wave turbulence seems to be the 
zero-flux thermodynamic solution which corresponds to equipartition of
energy in the $k$-space. In this case the theoretical solution 
derived in (Galtier et al., 2000) gives 
\begin{eqnarray}
E \sim k_\perp . 
\end{eqnarray}
We see that the scaling found numerically is somewhat flatter, i.e. $k_\perp^{0.6}$. The difference 
may be attributed, in particular, to the limited time of computation which does not allow to reach 
exactly the zero-flux solution. The problem is that the evolution slows down drammatically as one
moves closer to the lowest wavenumbers, and it takes an incredibly long computational time
to reach the final state.

\subsection{Global behavior}

\begin{figure}
\centerline{\psfig{figure=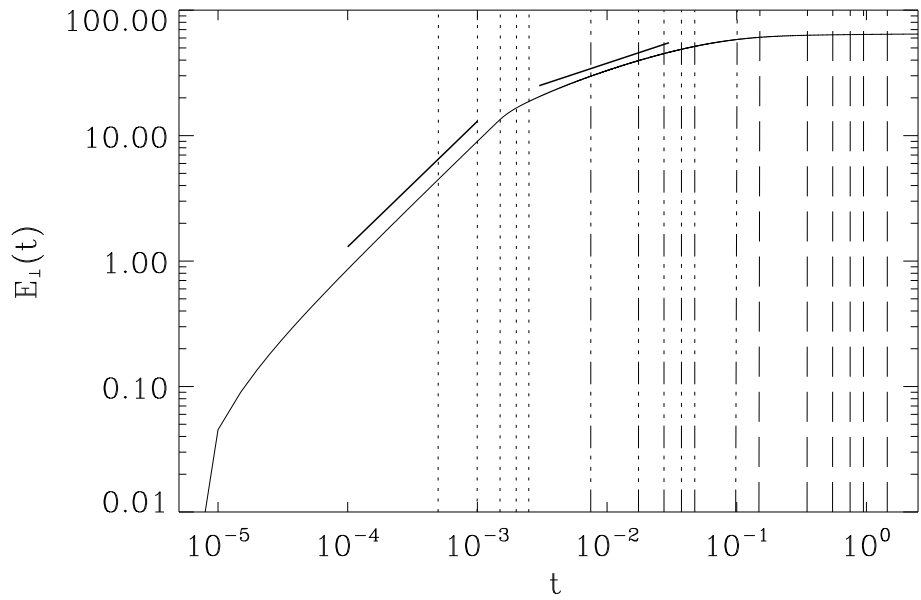}} 
\caption{Temporal evolution of shear-Alfv\'en wave energy for times up to $t_{18}$. For comparison, 
two slopes are given with power law indices $+1$ and $+1/3$ respectively. The vertical lines 
correspond to times $t_1$ to $t_{17}$.}
\label{Fig9}
\end{figure}

To complete our analysis, the temporal evolution of total shear-Alfv\'en wave energy is displayed in 
Figure \ref{Fig9} for times up to $t=2.5$. Different time scalings are clearly visible: a first power 
law in $t$ is found and a 
second in about $t^{1/3}$ is estimated. The first temporal power law is linked to the formation 
of the Kolmogorov-Zakharov solution during which a transient solution with spectrum $k_\perp^{-7/3}$ is 
produced (Galtier et al., 2000). The second temporal power law corresponds to the formation of the steady 
spectrum solution at intermediate scales discussed above. Finally, after time $t>0.2$ a plateau 
is reached with negligible energy variation in time (at least in logarithmic coordinate).

\section{Conclusion}

We have seen that the large-scale magnetic field is produced on a time-scale which is about 
$10^4$ larger than the time-scale needed to establish the small-scale energy cascade solution.
We believe that the final distribution in the large-scale part should correspond
to the thermodynamic energy equipartition $k_\perp^{1}$ even though it would take
an extremely long computing time for formation of this spectrum which we were
not able to achieve. Instead, we observed formation of a shallower long-term scaling
at the large scales, $k_\perp^{0.6}$, which we believe to be transient.
  
Because of the energy cascade to the smallest dissipative (resistive and viscous) scales,
the small-scale part of the spectrum is important for understanding the total
energy dissipation rate.
On the other hand, it is this large-scale part of the spectrum that contains most
of the wave (magnetic and kinetic) energy itself. One can view it as a sort of powerful
energy storage which is charged extremely slowly and in a very inefficient way (because
most of the charging energy is wasted via the energy cascade).
For example, if the forcing is concentrated in a thin spherical shell
with wavenumber lengths between $k$ and $k + \Delta k$
then the final energy at large scales (assuming the energy equipartition) will
be $k/\Delta k$ times greater than the energy at the forcing scales.

The mechanism of generation of large-scale magnetic
fields via nonlinear transfer from an energy source at smaller scales
may be relevant, in a very qualitative sense, to maintenance
of large-scale galactic fields by small-scale sources provided
by supernovae events. To pursue this line further, however, one would
have to consider a more realistic thin disk geometry, instead
of a simple homogeneous external field considered in the present paper,
which could be an interesting future project.

\section*{Acknowledgments}
Grants from the PNST (Programme National Soleil--Terre) program of INSU (CNRS)
are gratefully acknowledged.


\label{lastpage}

\begin{thebibliography}{}

\bibitem{Goldstein99}
M.L. Goldstein and D.A. Roberts, 1999, Magnetohydrodynamics turbulence in the solar wind.
{\itshape Phys. Plasmas} {\bfseries 6}, 4154--4160. 

\bibitem{galtier06}
S. Galtier, 2006, Multi-Scale Turbulence in the Inner Solar Wind. 
{\itshape J. Low Temp. Phys.} {\bfseries 145}, 59--74.

\bibitem{elmegreen}
B.G. Elmegreen and J. Scalo, 2004, Interstellar turbulence I: observations and processes.
{\itshape Annu. Rev. Astron. Astrophys.} {\bfseries 42}, 211--273. 

\bibitem{scalo}
J. Scalo and B.G. Elmegreen, 2004, Interstellar turbulence II: implications and effects.
{\itshape Annu. Rev. Astron. Astrophys.} {\bfseries 42}, 275--316. 

\bibitem{Govoni}
F. Govoni, M. Murgia, L. Feretti, G. Giovanni, K. Dolag and G.B. Taylor, 2006,
The intracluster magnetic field power spectrum in Abell 2255.
{\itshape Astron. \& Astrophys.} {\bfseries 460}, 425--438. 

\bibitem{pouquet}
A. Pouquet, 1993, Magnetohydrodynamic turbulence. 
In {\itshape Astrophysical fluid dynamics} (ed. J.-P. Zahn and J. Zinn-Justin). 
Elsevier science publishers, 139--227. 

\bibitem{widrow}
L.M. Widrow, 2002, 
Origin of galactic and extragalactic magnetic fields. 
{\itshape Rev. Mod. Phys.} {\bfseries 74}, 775--823. 

\bibitem{branden}
A. Brandenburg and K. Subramanian, 2005,
Astrophysical magnetic fields and nonlinear dynamo theory.
{\itshape Phys. Reports} {\bfseries 417}, 1--209. 

\bibitem{ferriere}
J.L. Han, K. Ferriere and R.N. Manchester, 2004,
The spatial energy spectrum of magnetic fields in our galaxy.
{\itshape Astrophys. J.} {\bfseries 610}, 820-826. 

\bibitem{galt00}
S. Galtier, S.V. Nazarenko, A.C. Newell and A. Pouquet, 2000, 
A Weak Turbulence Theory for Incompressible MHD. 
{\itshape J. Plasma Phys.}{\bfseries 63}, 447--488.

\bibitem{galt02}
S. Galtier, S.V. Nazarenko, A.C. Newell and A. Pouquet, 2002,
Anisotropic Turbulence of Shear-Alfv\'en Waves. 
{\itshape Astrophys. J.} {\bfseries 564}, L49-L52. 

\bibitem{naza01}
S.V. Nazarenko, A.C. Newell and S. Galtier, 2001,
Nonlocal MHD Turbulence. 
{\itshape Physica D} {\bfseries 152-153}, 646-652. 

\bibitem{davidson}
P.A. Davidson, 2004, Turbulence. 
Oxford University Press, New York. 

\bibitem{bigot}
B. Bigot, S. Galtier and H. Politano, 2007, 
Energy decay laws in strongly anisotropic MHD turbulence.
Submitted to {\itshape Phys. Rev. Lett.}

\end{thebibliography}
\end{document}